
\input harvmac
\let\norefs=n
\def\MTitle#1#2#3{\nopagenumbers\abstractfont\hsize=\hstitle\rightline{#1}%
\vskip 1in\centerline{\titlefont #2}\vskip .08in\centerline{\titlefont #3}
\abstractfont\vskip .5in\pageno=0}
\font\zfont = cmss10
\def\ZZ{\hbox{\zfont Z\kern-.4emZ}}
\def\qslsh{/\kern-.5em q}
\def\parslsh{/\kern-.5em \partial}
\def\gsim{{~\raise.15em\hbox{$>$}\kern-.85em
          \lower.35em\hbox{$\sim$}~}}
\def\lsim{{~\raise.15em\hbox{$<$}\kern-.85em
          \lower.35em\hbox{$\sim$}~}}
\def\la#1{\lambda_{#1}}
\def\ru#1{r^{(1)}_{\tilde {#1}}}
\def\rd#1{r^{(2)}_{\tilde {#1}}}
\def\p#1{p_{\tilde {#1}}}
\def\Tu#1{T^{(1)}_{#1}}
\def\Td#1{T^{(2)}_{#1}}
\def\tT#1{\tilde T_{\tilde  {#1}}}
\def\etab{\bar\eta}
\def\Fi0d{\Phi_0^{\dagger}}

\def\fil{\varphi}
\def\fild{\varphi^{\dagger}}
\def\G{\Gamma}
\def\Gd{{\Gamma^{\dagger}}}
\def\tG{\tilde{\Gamma}}
\def\tGd{{\tilde{\Gamma}}^{\dagger}}
\def\tZ{\tilde Z}

\def\ie{{\it i.e.,}}

\def\c#1{\epsilon_{#1}}
\def\c#1{\epsilon_{#1}}
\def\c#1{\epsilon_{#1}}
\def\gws{SU(3)\times SU(2) \times U(1)_Y}

\MTitle{hep-ph/9507315, RU-95-45}
{A Note on the Effective Soft SUSY-Breaking}
{Lagrangian Below the GUT Scale}

\bigskip
\centerline{Riccardo Rattazzi
\foot{E-mail: rattazzi@physics.rutgers.edu}}
\smallskip
\centerline{\it Department of Physics and Astronomy}
\centerline{\it Rutgers University}
\centerline{\it Piscataway, NJ 08855-0849, USA}
\bigskip
\noindent
I consider the superfield derivation of the
effective theory of softly broken supersymmetry
below the GUT scale. I point out the role of supergauge invariance
in determining the form of the result, which is rather restricted
in interesting classes of models. As an example I discuss sfermion
mass splittings for matter embedded in a single GUT
multiplet. Interesting differences arise between the cases of $SU(5)$,
$SO(10)$ and $E_6$.


\Date{}

\lref\dinenel{M. Dine and A.E. Nelson,
{\it Phys. Rev.} {\bf D48} (1993) 1277 (hep-ph/9303230);
Santa Cruz preprint SCIPP-94-21, (hep-ph/9408384).}
\lref\girard{L. Girardello and M.T. Grisaru, {\it Nucl.
Phys.} {\bf B194} (1982) 65.}
\lref\sohn{M.F. Sohnius, {\it Phys. Rep.} {\bf 128} (1985) 39.}
\lref\pom{A. Pomarol and S. Dimopoulos, CERN-TH/95-114, hep-ph
9505302, May 1995.}
\lref\hitoshi{Y. Kawamura, H. Murayama and M. Yamaguchi, {\it
Phys. Rev.} {\bf D51} (1995) 1337.}
\lref\hall{L. Hall, J. Lykken and S. Weinberg,
 {\it Phys. Rev.} {\bf D10} (1983) 2359.}
\lref\fayet{P. Fayet, {\it Nuovo Cim.} {\bf 31A} (1976) 626.}
\lref\anderson{G. Anderson {\it et al.,} {\it Phys. Rev.} {\bf D49}
(1994) 3660.}
\lref\anantha{B. Ananthanarayan, G. Lazarides and Q. Shafi, {\it
Phys. Rev.} {\bf D44} (1991) 1613;
L.J. Hall, R. Rattazzi and U. Sarid, {\it Phys. Rev.}
{\bf D50} (1994) 7048.}
\lref\rs{R. Rattazzi and U. Sarid, SU-ITP-94-16, RU-95-13,
hep-ph/9505428, May 1995.}
\lref\soni{S.K. Soni and H.A. Weldon, {\it Phys. Lett.} {\bf 126B}
(1983) 215.}
\lref\louis{ V.S. Kaplunovsky and J. Louis, {\it Phys. Lett.}
{\bf 306B} (1993) 268.}
\lref\sen{A. Sen, {\it Phys. Rev.}, {\bf D30} (1984) 2608;
 {\it ibid.,} {\bf D32} (1985) 411.}
\lref\barbieri{R. Barbieri, G. Dvali and A. Strumia, {\it Nucl. Phys.}
{\bf B391} (1993) 487}
\lref\missing{S. Dimopoulos and F. Wilczek, ITP Santa Barbara Preprint
UM HE 81-71 (1981) and Proceedings Erice Summer School,
Ed. A. Zichichi (1981).}
\lref\babu{K.S. Babu and S. Barr, {\it Phys. Rev.}
{\bf 48D} (1993) 5354}
\lref\raby{L.J. Hall and S. Raby, OHSTPY-HEP-T-94-023,
hep-ph/9501298, Jan. 1995}
\lref\giudice{G.F. Giudice and E. Roulet, {\it Phys. Lett.}
 {\bf 315B}
(1993) 107.}
\lref\lizu{Z. Berezhiani, C. Cs\`aki and L. Randall, INFN-FE 01-95,
MIT-CTP-2404, hep-ph/9501336.}
\lref\anselm{A. Anselm and A. Johansen,
{\it Phys. Lett.} {\bf 200B} (1988) 331.}
\lref\dvali{Z. Berezhiani and G. Dvali, {\it Sov. Phys. Lebedev
Inst. Rep.} {\bf 5} (1989) 49.}
\lref\bdm{R. Barbieri, G. Dvali and M. Moretti, {\it Phys. Lett.} {\bf
312B} (1993) 137.}
\vfill
\eject

\noindent{\bf 1.}{\hskip 0.2truecm}In a recent interesting paper
Pomarol and Dimopoulos \pom\ used
the superfield formalism to derive the effective lagrangian
of softly broken supersymmetry below the GUT scale. That paper shows
that the use of superfields allows a great simplification with
respect to the same calculations performed in component notation,
like for instance Ref. \hitoshi\hall. In this note we will elaborate
on the method of Ref. \pom\ and show
a parametrization of the heavy superfields that simplifies the
calculation even further.
In our derivation it is
straightforward to ``power count'' and keep only the relevant effects,
and, more importantly, the origin of certain cancellations,
which were missed in Ref. \pom, is made clear. The result leads
to fairly constrained soft mass splittings in the
interesting class of models
where susy-breaking  takes place in a {\it hidden sector}.
The only  mass
splitting between light sparticles belonging to the same GUT irreps are
mediated by the heavy gauge fields. They generally consist of just
the so-called D-terms \hitoshi, though in particular cases additional
interesting effects may arise.

 This note is organized as follows. In this section
we focus on the most general soft breaking terms from a {\it hidden
sector}, \ie\ also allowing for non-flat Kahler metrics,
and described in equation (3) below.
In section 2 we discuss sparticle splittings respectively in the case
of $SU(5)$, $SO(10)$ and $E_6$ unification, including the scenario of
$SO(10)$ Yukawa unification \anantha. We point out the possibility
of important splitting effects, which had not been discussed before,
and which arise when the unified group in enlarged to $E_6$.
We also comment on
 case in which the MSSM
Higgs doublets are pseudo-Goldstones of ``accidental'' symmetries
 \dvali\bdm\lizu.
In section 3 we discuss  the
case of general soft terms.
In sect. 4 we conclude.

In what follows we describe our assumptions for the observable sector.
We consider a supersymmetric
Grand-Unified Theory with gauge group $G$ and with a set of chiral
matter fields $\Psi$. It is assumed that, in the absence of
 soft supersymmetry breaking terms, the v.e.v. $\Psi_0$ of the chiral
superfields breaks $G$ down to a subgroup $H$, while keeping
supersymmetry unbroken. We indicate the set of vector superfields by
$V=(V_A,V_a)$ where $V_A$ ($A,B,\dots$) are the massive ones
corresponding  to the broken generators
$T_A$, while $V_a$ ($a,b,\dots$) correspond to the unbroken
generators. The set of chiral superfields can be decomposed as
$\Psi=(\Phi_A,\Phi_k,\fil_\alpha)$, where $\Phi_A$ ($A,B,\dots$) are
the Goldstones eaten via the super-Higgs mechanism, $\Phi_k$
($k,l,\dots$) are heavy non-Goldstone fields, and finally
$\fil_\alpha$ ($\alpha,\beta,\dots$) are the massless non-Goldstone
fields corresponding to the matter content of the low-energy effective
theory. We choose a basis for the broken generators $T_A$ such that the
vectors  $e_A=T_A\Psi_0/|T_A\Psi_0|$ form an orthonormal basis of the
Goldstone subspace, \ie\  $e_A^\dagger e_B=\delta_{AB}$.
The v.e.v.'s of the various fields are indicated by a ``0''
sub- or supscript. Unbroken supersymmetry implies $\Psi_0^\dagger T_A
\Psi_0=0$, which in component notation reads $\Phi_A^0=0$.
Notice though that in general $\Phi_k^0\sim M_G\not = 0$.
We will assume that there are no light fields that are siglets
under both $H$ and any (possibly discrete) low-energy global
symmetries; then there are no GUT v.e.v.'s associated with the light
fields, \ie\  $\fil_\alpha^0=0$ and also no terms linear in
$\fil_\alpha$ can appear in the low energy lagrangian.
In the
chosen $\{A\}$ basis the mass matrix of the heavy vectors is
 diagonal and given by $M^2_{AB}=\Psi_0^\dagger \{T_A,T_B\}\Psi_0=
M_A^2 \delta_{AB}$ and $M_A={\sqrt 2}|T_A\Psi_0|$.
(Notice that since we work at zero momentum we can take the gauge
coupling $g=1$).
The degrees of freedom represented by $\Phi_A$ are superfluous and,
as done in Ref. \pom, can be eliminated by going to the super-unitary
gauge \fayet
\eqn\gauge {  \Phi_A\propto {\Psi^\dagger}_0 T_A \Psi=0}
By going to this gauge, the superpotential is written in terms of the
light fields $\fil_\alpha$ and shifted heavy $\psi_k=\Phi_k-\Phi_k^0$
as
\eqn\superpot{W={1\over 2}\mu_{kl}\psi_k\psi_l +{1\over 2}\la{k\alpha\beta}
\psi_k\fil_\alpha\fil_\beta+{1\over
2}\la{kl\alpha}\psi_k\psi_l\fil_\alpha+{1\over
3!}\la{klm}\psi_k\psi_l\psi_m+\tilde W(\fil)}
where $\mu_{kl}$ are masses of order $M_G\simeq 10^{16}$ GeV,
$\lambda$'s are Yukawa couplings and
$\tilde W$ is a piece which depends on the light fields only.
Let us now introduce the soft susy breaking terms.
We will write them in terms of the spurion $\eta=m\theta^2$ where
$m\sim m_Z$ \girard.
We also follow the conventions of Ref. \sohn, where $d^2\theta=D^2/2$ and
$\int d^2 \theta \theta^2=-2$.
As discussed in Ref. \hitoshi, the most general lagrangian, inclusive
of soft breaking terms, can be written in {\it hidden sector}\foot{
Our definition of {\it hidden sector} models corresponds to the
situation in which, in the parametrization of Ref. \soni, the
superpotential of the original supergravity
splits into the sum of two pieces, one depending
only on the fields of the susy-breaking sector and the other only
on the observable fields. The form (3) is also stable in perturbation
theory \hitoshi\barbieri.}  scenarios as
\eqn\lagrange{\eqalign{
{\cal L}=&{1\over 4}\int d^4\theta \Bigl\{\Psi^\dagger (1+\etab \Gd)
e^{2V}(1+\eta \G)\Psi + \Psi^\dagger e^{2V}\tZ\eta\etab\Psi+
 (\Psi^T \Lambda_1 \Psi +{\rm h.c.})\eta\etab\cr
+&(\Psi^T\Lambda_2\Psi\etab +{\rm h.c.})\Bigr \}
-{1\over 2}\left \{\int d^2\theta (1+a\eta)W+{\rm h.c.}\right \}\cr}}
where $\G$, $\tZ$, $\Lambda_1$ and $\Lambda_2$ are $G$-invariant matrices
while $a$ is just a c-number. In the chosen gauge $\Psi$ is
represented by $\Psi=\Psi_0+\psi+\fil$, in an obvious vector notation.
We are following the notation of Ref. \pom, apart from having
``factored'' the $\G$'s, \ie\ our $\tZ$ corresponds to $Z-\Gd\G$ of that
paper. We want now to integrate out the heavy fields and take the
double limit $m={\rm const.}$, $m/M_G\to 0$. The low energy effective
lagrangian is completely determined by its form at $V_a=0$ (we remind
that the $V_a$ correspond to the unbroken generators $T_a$). Indeed,
when $H$ contains an abelian factor $Y$, consistently with
the
low energy gauge invariance, one might expect a Fayet-Iliopoulos
D-term proportional to $V_Y$. It is however clear that, from
eq. \lagrange\ at $\fil=0$,
this term is bound to be proportional to $Y \Psi_0$ which
is equal to $0$. (Notice however that by integrating out the heavy
modes at 1-loop we would get such a term, proportional to ${\rm
Tr}Y\hat m^2$ where $\hat m^2$ is the soft contribution to the heavy masses).
 Thus we can derive the effective lagrangian for
$V_a=0$, and then obtain the complete one by covariantizing in $H$
the existing interactions, \ie\ just by inserting $\exp(2V_aT_a)$ in the
Kahler potential.
The relevant effective interactions, \ie\ those that survive the above
limit, are suitably characterized by
making the following classification of infinitesimal quantities:
\eqn\infin{\eta={\cal O}(\epsilon_1)\quad\quad
\fil={\cal O}
(\epsilon_1\epsilon_2 )\quad \quad D^2={\cal O}(\c{1}\c{2}\c{3} )}
In what follows an operator is defined to be ${\cal O}(\c{1}^n)$ when,
according to \infin, it involves n-powers of $\c{1}$, and analogously
for $\c{2}$, $\c{3}$.
Then in the assumption that $\fil$ does not contain any singlet
of the full (gauge + global) low energy symmetry, we have that
the relevant low energy lagrangian is written as $\int d^4\theta
K_{eff} +(\int d^2\theta W_{eff}+{\rm h.c.})$ where
\eqn\relevant{K_{eff}\geq {\cal O}(\epsilon_1^4),\,{\cal O}(\epsilon_2^2),\,
{\cal O}(\epsilon_3^0)\quad\quad
W_{eff}\geq {\cal O}(\epsilon_1^5),\,{\cal O}(\epsilon_2^3),\,
{\cal O}(\epsilon_3).}
(For instance, $K_{eff}$ is a polynomial of order $\leq
4$ in $\eta,\fil,D^2$, but of order $\leq 2$ in $\fil,D^2$ and in
fact of order $0$ in $D^2$, since there are no singlets.)
To satisfiy the above relations we  just need to solve the
equations of motions for $V_A$ and $\psi_k$ to a finite order in the
$\epsilon$'s. One important remark here is that the solutions $V$ and $\psi$
to the equations of motion are at least ${\cal O}(\c{1})$.
 To proceed we need to expand eq. \lagrange\ in a power series in $V$
and $\psi$. Before doing so it is useful to perform the following
field redefinition
\eqn\redefpsi{\psi_k\to
\psi_k-\mu_{kl}^{-1}\la{l\alpha\beta}\fil_{\alpha}\fil_{\beta}/2}
which essentially takes care of the lowest order equations of motion
for $\psi$. In this new parametrization the $V,\psi$ dependent part of
eq. \lagrange\ has the form
\eqn\lform{\eqalign{&\int d^4\theta \left \{V^2+V(\eta+{\cal
O}(\epsilon_1^2)+\psi{\cal O}(\epsilon_1)+{\rm
h.c.})+V^3+V^2\eta+\psi\etab+\psi\eta\etab+\dots\right\}\cr
&+\left\{\int d^2\theta^2 \bigl [\psi^2+ (\psi-\fil^2)^2\fil+
(\psi-\fil^2)^3\bigr ](1+a\eta)+{\rm h.c.}\right\}\cr}}
where we took the GUT masses to be ${\cal O}(1)$. Notice that the
simple terms $V$ and $V(\psi+\fil)$ are absent respectively because of
zeroth-order unbroken susy and because $\psi$ and $\fil$ are
orthogonal to the Goldstone states.
By the dots we mean even higher orders in $V$ and $\psi$.
The equations of motion then have the form
\eqn\motionpsi{\psi +\fil^3 +\bar D^2{\cal
O}(\c{1})+\psi\fil+(\psi-\fil^2)^2=0.}
and
\eqn\motionv{V+(\eta+{\cal
O}(\epsilon_1^2)+\psi{\cal O}(\epsilon_1)+{\rm
h.c.})+V^2+V\eta+\dots=0}
Where in eq. \motionpsi\ the $D^2$ term comes from varying the Kahler
potential.
The solution is then $\psi\sim \c{2}^3+\bar{D}^2\epsilon_1$, which
manifestly does not give any relevant effects both in $W$ and in
$K$ (cfr. eqs. \infin\relevant\ and \lform\ above). In the case of
$V$ the situation seems
more complicated since the solution is $V\sim\c{1}+\c{1}^2+\dots$, and also
the higher order terms $V^3$ and $V^2\eta$ contribute to the relevant
terms. This would not happen if the term $V\eta$ in \lform\ were
missing, so that
$V\sim\c{1}^2$. In this situation the only relevant pieces would arise
from $V^2+V\c{1}^2$ by a trivial quadratic integration.
In fact eq. \lagrange\ possesses a reparametrization invariance,
related to the gauge symmetry, and by means of which the $V\eta$
term can be eliminated. Consider the following gauge-type
field redefinition
\eqn\vredef {e^{2V}\,\to\, (1-c_A^* T_A\etab)e^{2V}(1-c_A T_A\eta).}
It is equivalent to
\eqn\gredef{\G\,\to\,\G-c_A T_A=\tG}
in the full lagrangian, since eq. \vredef, being like a gauge
transformation, does not affect the gauge kinetic term. Then eq. \gredef\
defines a reparametrization invariance which turns out to be
very useful. The $V\eta$ term in
eq. \lform\ (see eq.\lagrange) is in fact $\propto \Fi0d T_A\tG\Phi_0$,
which is $=0$ if we choose
\eqn\nolinear{c_A=2\Fi0d T_A\G\Phi_0/M_A^2}
in eq. \gredef. In what follows $\tG$ is defined by eqs. \gredef\nolinear.
 Notice that, when $c_A\not = 0$, $\tG$ no longer
commutes with $G$, but $[T_A,\tG]$ is a gauge generator. Then from
unbroken susy and the gauge condition \gauge , we have that both
$\Fi0d [T_A,\tG]\Phi_0$ and $\Fi0d [T_A,\tG]\Psi$ vanish.
 Notice also that, by the Wigner-Eckart theorem, $c_A =
0$ when $T_A$ is not a singlet of $H$, so that the above field
redefinition is important only when ${\rm rank}(H)<{\rm rank}(G)$.
Then, by  \nolinear , $c_AT_A$ is an element of the broken
Cartan subalgebra. For instance, in $SO(10)$ we have only $c_AT_A\propto
X$, where $X$ is the broken Cartan generator which is orthogonal
to the hypercharge $Y$, \ie\ $SO(10)\supset SU(5)\times U(1)_X$.
As we said, by eqs. \gredef\ and \nolinear\ the solution
is $V_A={\cal O}(\c{1}^2)$ and the
$V$-dependent part in eq. \lagrange\ reads
\eqn\vquadr{{1\over 4}\int d^4\theta \left\{M_A^2 V_A^2+2V_A\bigl [
M_A^2 d_A\eta\etab +\fild T_A \fil +(\Fi0d T_A\tG\fil\eta+\fild
T_A\tG\Phi_0\eta + {\rm h.c.})+{\cal O}(\c{1}^3)\bigr ]\right \}}
where $d_A=\Fi0d(\tGd T_A \tG +T_A \tilde Z)\Phi_0/M_A^2$. Then,
the integral in $V_A$ is trivial and gives
\eqn\softgauge{-{1\over 2}\int d^4\theta \left \{\fild T_A\fil
{d_A}+ {|\Fi0d T_A\tG\fil+\fild T_A\tG \Phi_0|^2\over
M_A^2}\right \}\eta\etab+{\cal O}(\c{1}^5).}
Thus the effective lagrangian for the light fields is given by
eq. \softgauge\ plus the following
\eqn\softrest{\eqalign{
{1\over 4}\int d^4\theta \Bigl \{& \fild (1+\etab \tGd)e^{(2V_aT_a)}(
1+\eta \tG)\fil +\fild e^{2V_a T_a} \tilde Z\eta\etab\fil +[
\fil ^T(\Lambda_1\eta+\Lambda_2)\fil\etab
+{\rm h.c.}]\cr
&-\bigl[{1\over 2}\Fi0d(\tGd\etab +\tGd\tG\eta\etab +\tilde Z\eta\etab)
+\etab\Phi_0^T(\Lambda_1\eta+\Lambda_2))\bigr ]_k
\mu_{kl}^{-1}\la{l\alpha\beta}\fil_\alpha\fil_\beta+{\rm h.c.}\Bigr
\}\cr
&-{1\over2}\left \{ \int d^2\theta^2 (1+a\eta)\tilde W(\fil) +{\rm
h.c.}\right \}\cr}}
where the terms in the second line arise from the $\psi$ redefinition
described in eq. \redefpsi. We stress that the use of the
reparametrization \gredef\ toghether with eq. \nolinear\ leads to a
great simplification. For instance, by using Feynman diagrams as in
Ref. \pom, we would have just two diagrams from the gauge sector, compared
to the eleven of Ref. \pom.
\vskip0.5truecm
\noindent{\bf 2.}{\hskip 0.2truecm} Let us focus  on the
 chirality preserving
soft masses induced by the above. Defining, to match the notation
of Ref. \pom , $\tilde Z=Z-\Gd\G$
we can write the result as
\eqn\quadratic{\eqalign{{1\over 4}\int d^4\theta &\Bigl \{\bigl
[\fild(Z-2T_Ad_A)\fil +(|\tG_{k\alpha}\fil_{\alpha}|^2- |\G_{k\alpha}
\fil_{\alpha}|^2)-(|\tGd_{A\alpha}\fil_{\alpha}|^2+|\G_{A\alpha}
\fil_{\alpha}|^2)\bigr ]\eta\etab\cr
-&|\G_{\beta\alpha}\fil_\alpha|^2\eta\etab+
[\fild(1+\etab\tGd)]_\alpha [(1+\eta\tG)\fil]_\alpha\Bigr\}\cr}}
where we have used $M_A={\sqrt 2}|T_A\Phi_0|$ in eq. \softgauge, and
summation over $k,\,\beta,\, A$ is also understood.
Notice that the second term in the second line is in fact giving no
contribution to chiral preserving masses, since it is reduced to
$\fild\fil$,
by the field redefinition $\fil_\alpha\to [(1-\eta\tG)\fil]_\alpha$,
which leaves only A- and B-type soft terms from the superpotential.
The form of eq. \quadratic\ is then fairly restricted. It is
remarkable that the superpotential couplings do not enter directly
in the above equation, though they obviously affect the $\Gamma$'s and
$Z$'s via RG evolution. Notice, indeed,
that for the subset of the light fields $\fil_\alpha$ upon
which the broken Cartan subalgebra acts diagonally (as it happens
to ordinary matter in most $SO(10)$ models), we have that
$\tG_{k\alpha}=\G_{k\alpha}$ and $\tGd_{A\alpha}=\Gd_{A\alpha}$ so that
the second term in brakets in \quadratic\ vanishes. It is interesting
to study the splittings of light sfermions embedded  within a single
GUT multiplet. In the cases of interest, the broken Cartan subalgebra
acts diagonally on these fields. Then we are reduced to just consider
D-terms and $\G_{A \alpha}$, $\G_{\alpha A}$.
 Notice though that
only the D-terms are of gauge nature, which is to say universal
(not with the meaning of ``degenerate''!)
and generally expected,
while $\G_{A\alpha}$ and $\G_{\alpha A}$ correspond
to  mixings between light
fields and Higgs multiplets, or more precisely between light fields
and heavy vector superfields. Their potential interest then turns out
to depend strongly on the original gauge group $G$. In what follows
we will consider their effect on matter sfermions in the MSSM
respectively for $G=SU(5)$, $SO(10)$ and $E_6$. In $SU(5)$ the massive
vectors transform under $G_{WS}=\gws$ as $(\bar 3, 2, -7/6)$ plus its
conjugate. Then there is no mixing to the MSSM fields, and the
sparticles from the same $\underline {10}$ and
$\underline {\bar 5}$ are unsplit. When $G=SO(10)$ there is an
additional set of heavy vectors tranforming under $G_{WS}$ as
$\underline {10}\oplus\underline{\bar{10}}$ of $SU(5)$. These can in
principle mix with the matter fields $Q,U_c$ and $E_c$. The
mixings are induced by the terms $\int d^4\theta \underline
{16}_H^\dagger (\alpha_i \eta+\beta_i\etab)\underline {16}_i$, where
$\langle \underline {16}_H\rangle\not = 0$ and $\underline {16}_i$,
$i=1,2,3$ contain the three light families. However the
corresponding $\G_{A\alpha}$, $\G_{\alpha A}$ break $R$-parity,
(i.e. there is no way to define an unbroken $R$-parity under which the
heavy vectors associated with $SO(10)\to SU(5)\times U(1)$ are odd),
so that we expect them to be rather
small or absent at all.
In fact they lead to the $R$-odd terms $H_u L_i$ (both
supersymmetric and soft) from the second line of eq. \quadratic.
These are generated by integrating out the right-handed neutrinos.
Notice that for particular choices of the soft terms or of the
neutrino mass matrix (like when the neutrinos $N_i$ get Dirac masses by
mixing to matter singlets $S_i$) these  $R$-odd masses could be
absent at tree level, but they would still be  generated at
1-loop. Thus we conclude that for sfermions within the same
$\underline {16}$ of $SO(10)$ the only relevant source of mass
splitting is given by the D-term associated with the only broken
Cartan generator. The situation can be fairly different in $E_6$.
When $E_6 \to G_{WS}$, with respect to the previous case,
 there are additional heavy vectors
in the $\underline {16}\oplus\underline {\bar{16}}\oplus \underline 1$
of $SO(10)$.  Now, we
can endow the theory with $R$-parity in such a way that the
vectors in $\underline {16}\oplus \underline {\bar {16}}$
are $R$-odd, thereby allowing their soft mixing with
matter.  These are induced by soft terms involving
$\underline {27}_H^\dagger \underline {27}_M$, where $\langle
\underline {27}_M\rangle \not = 0$ and matter is contained in
$\underline {16}_M\subset \underline {27}_M$ (with obvious notation).
 The crucial
remark here is that if $\langle \underline {27}_H
\rangle$ breaks $E_6$ down to $SO(10)\times U(1)$, the $\G_{A \alpha}$
between matter and the vector $\underline {16}$'s are non zero, and still
we have the $\ZZ_4$ center of $SO(10)$ unbroken. It is then possible that
its $\ZZ_2$ subgroup $Z$, or a combination of it with a global one,
survives and corresponds to
$R$-parity. A necessary condition for this to happen is that
 all $E_6$ Higgs multiplets
get vacuum expectation values along directions with definite $Z$. In
this case a  combination of $Z$ and other
global discrete symmetries could be
the low energy R-parity. For instance the rank could be further
reduced by the $Z$-odd vev of a $\underline {27'}_H$. In this case there
should be a global parity $Z'$ under which $\underline {27'}_H$ is odd
and $R=ZZ'$. Now, the main point is that when $E_6\to \gws$ directly
at $M_G$, the masses of the $R$-odd vectors in the $\underline {16}$
are in general all split by ${\cal O}(1)$. In fact this fields will also
get mass from $SU(5)$ breaking vevs like, for instance, a $\langle
\underline {78}_H\rangle$.
As a result the sfermion masses induced by eq. \softgauge\ will
be clearly respecting only the low-energy gauge symmetry.
While it may not be  easy
to obtain a model with these features, it is an amusing fact that
by enlarging the gauge group the symmetry properties of the
soft masses are in principle reduced

We are thus lead to the interesting conclusion that, unless $R$-odd
gauge bosons appear,
 for matter belonging
to well definite GUT irreps, the only sources of soft masses that we
expect to be important are
the $G$ invariant ones plus a $D$-term for each broken Cartan
subalgebra generator. However when $G\supset E_6$, there can be $R$-odd
vectors and a new class of contributions is allowed.
 When  the MSSM Higgs  doublets $H_{u,d}$ sit
in specific GUT irreps
also this sector is fairly constrained. This is indeed what happens
for the interesting $SO(10)$ Yukawa unified situation
\anantha\ in which both light Higgses lie in the same
 $\underline{10}$. In this case the $\G_{A\alpha}$ terms
 in \quadratic\ is also vanishing for the Higgs doublets, since
$\underline {10}$'s have zero  v.e.v.. Then, neglecting
$R$-parity violating terms, the soft masses of the third sfermion
family + Higgses are completely specified by the three parameters
$m_{10}^2$, $m_{16}^2$ and the  D-term $d_X$. It has been shown
 in Ref. \rs\ that this constrained form of the soft terms leads
 to difficulties
in radiative electroweak breaking.  A more
plausible picture of radiative electroweak symmetry breaking, requires
additional splittings. As shown above these can arise when SO(10) is
enlarged to $E_6$. In sect. 3, we show that they can also arise in
$SO(10)$, but by allowing very general soft terms.


Notice that in Ref. \pom\ some important cancellations leading to eq.
\quadratic\  were overlooked, so that other contributions of
genuine gauge type in
addition to the D-terms were claimed.  Indeed these
cancellations, in the computation of Ref. \pom , arise from the
form of the lowest order solution $V\sim\eta+\etab$,
which is a pure gauge configuration, and are manifest by using eq. \gredef.
Notice that eq. \quadratic\  agrees with the result in Ref. \hitoshi.
Indeed, we have compared our full effective lagrangian with the
complete result
given in eqs. (3.47), (3.48) of Ref. \hitoshi. We found agreement
for  all terms apart from the quadratic (chiral-breaking) B-type
ones.
For these terms there is a mismatch proportional to the $c_A$'s.
 However, this is
probably due to a typographical error, since the missing terms are
included in  eq. (3.49) of the same Ref., which displays B-type masses only.

Notice that in most cases  of
interest, it is $(c_BT_B)_{A\alpha} = 0$
so that $\tG_{A\alpha}=\G_{A\alpha}$.
However, one has in general $(c_B T_B)_{A\alpha}\not = 0$
in the interesting class of models where
the  MSSM Higgs doublets have the interpretation of pseudo-Goldstone bosons
of an accidental $G\times G$ symmetry of the Higgs sector of the
superpotential.
Models of this type have attracted attention \dvali\bdm\lizu\
as they offer an elegant solution to the doublet-triplet splitting problem.
We devote the remainder of this section  to briefly recall their properties
and to discuss the implications of \softgauge\softrest\ on the
pseudo-Golstone masses.
The Higgs superpotential is supposed to have the form
 $W_h=W_1(\Psi_1)+W_2(\Psi_2)$, where $\Psi_1$ and $\Psi_2$ are separate sets
of fields that transform non-trivially under $G$.
 Thus $W_h$ has a $G\times G$
symmetry, and we indicate respectively with $\Tu{}$ and
$\Td{}$ the $G$ generators acting on each sector (the gauge generators
are then given by $\Tu{A}+\Td{A}$).
When $\Psi_1^0$ and $\Psi_2^0$ indpendently preserve supersymmetry
(\ie\ the contribution to the gauge $D$-terms is zero in both sectors), and
the S.S.B. pattern is $G_1\to H_1$, $G_2\to H_2$ with $H=H_1\cap H_2$, there
is a doubling of the Goldstones belonging to the subspace $G/H_1 \cap G/H_2$.
\foot{Notice that, consistently with our general assumptions
we are limiting ourselves to the case in which $\Psi_{1,2}^0$ are determined
 before the introduction of soft breaking terms. In some realistic attempts,
however, like Ref. \bdm\ and model I in Ref. \lizu, this may not be the case.
Model II of \lizu, however, satisfies our assumptions.}
For each generator in this set, there is a ``gauge'' Goldstone eliminated
by the super-Higgs mechanism, but in addition there is a physical massless
chiral superfield. We can associate these  fields
to $G\times G$ generators $\tT{A}=\ru{A}\Tu{\tilde A}+\rd{A}\Td{\tilde
A}$ ($\tilde A,\,\tilde B,\dots$),
where $r^{1,2}_{\tilde A}$ are numerical coefficients
defined so that the vectors $\p{A} =\tT{A}\Psi_0$,
 have unit norm and are orthogonal to the gauge Goldstones,
\ie\ $e^\dagger p=0$.
For instance in \dvali\bdm\lizu\ $G=SU(6)$ and $H_1=SU(4)\times SU(2)\times
U(1)$, $H_2=SU(5)$ with $H=SU(3)\times SU(2)\times U(1)$,
so that the pseudo-Goldstones
are just the two doublets $H_u\oplus H_d$. The result is just
a consequence of the group algebra and of the split form of the
superpotential.
As already mentioned, it is  clear that $(c_AT_A)_{B \tilde C}\not =
0$, and the general result \softgauge \softrest\ has to be used,
in order to discuss the soft terms. To do so
we construct the pseudo-Goldstone superfields as
$\fil_g=\fil_{\tilde A}\p{A}$.
Then we notice that the $G\times G$ symmetry of $W_h$ implies
\eqn\symm{\mu_{kl}^{-1}\lambda_{l\tilde A\tilde B}=-(\tT{A}\tT{B}\Psi_0)_k=
-(\tT{B}\tT{A}\Psi_0)_k.}
{}From whence, remarkably,
the mass matrix \softgauge\softrest\ of the pseudo-Goldstones depends
very  little on the details
of the superpotential, the only parameters entering the definition of the
$\tilde T$'s being the two v.e.v.'s $\Psi_1^0$ and $\Psi_2^0$. It is
interesting
to consider the case in which $\Gamma$ $\tilde Z$ and $\Lambda_{1,2}$
are also
$G\times G$ invariant. Then from eq. \symm\ we immediately get that
the terms in
eq. \softrest\ proportional to $\Lambda_{1,2}$ vanish for the
 $\fil_{\tilde A}$.  By writing the scalar components as
$\fil_{\tilde A}=\sigma_{\tilde A}+i\pi_{\tilde A}$, where $\sigma$ and
$\pi$ are real scalars, it can be shown
through straightforward, though tedious, calculations that the mass
contributions from \softgauge\softrest\ are only of the form $\sigma^2$
and $\sigma\pi$, with no $\pi^2$ terms.\foot{
We assume that
$\fil_{\tilde A}$ contains no $H$ singlets. In this situation
$\Fi0d \tG T_A\Phi_0=0$ implies also $\Fi0d \tG T^{(1)}_A\Phi_0=\Fi0d
\tG T^{(2)}_A\Phi_0=0$, which is also of
considerable help in the computations.}
Thus there remain flat directions $\sigma=0$, $\pi\not =0$, which
correspond indeed to  genuine pseudo-Goldstones. \foot{ This could be
deduced by inspection of the full-potential
inclusive of soft terms, before integrating out the GUT fields. In
this  respect, the absence of the $\pi_{\tilde A}^2$ terms
constitutes a non trivial check of eqs. \softgauge\softrest.}
However, we stress that, in contrast with the
case of universal soft terms \giudice, there are in general mixing terms
$m^2_{\tilde A \tilde B}\sigma_{\tilde A}\pi_{\tilde B}$, so that the
mass matrix
is not definite semi-positive. The appearance of these mixing terms
is closely
related to the appearance of D-term type splittings. For instance, when
$\fil_g=H_u\oplus H_d$  we have
\eqn\flat{{\cal L}_{mass}=(H_u^*,H_d)\left( \matrix {m_0^2+\Delta^2 &
 m_0^2\cr
m_0^2 & m_0^2-\Delta^2\cr}\right )\left (\matrix{H_u\cr H_d^*\cr}\right )}
where $m_0^2$ and $\Delta^2$ are soft mass parameters.
Then the general signature of models of this type
is $m_u^2+m_d^2-2|B\mu|=0$ at the GUT scale.
Finally, another possibility
given by non-universal soft terms is to allow an explicit breaking
of the $G\times G$ symmetry by the $\Gamma$, $Z$ and $\Lambda$'s
themselves: in this case
a positive diagonal piece can be added to the above mass matrix, giving the
possibility to stabilize the GUT scale tree level potential at $H_{u,d}=0$.
In this case however the prediction \flat\ is lost.

\vskip 0.5truecm
\noindent{\bf 3.}{\hskip 0.2truecm}What we have done so far was limited
to the scenario in which, in the parametrization of Ref. \soni, the
source of supersymmetry breaking interacts with  observable matter
only in the Kahler potential. In more general scenarios, as discussed for
instance in Refs. \soni\ and \louis, there will also be susy-breaking
feed-down via superpotential couplings. In this situation we do not
expect for the soft terms the restricted form of eq. \lagrange.
In what follows
we just want to sketch how the integration of the heavy GUT modes would be
performed in the most general situation.

By a field redefinition we can write all the soft terms which are linear in
$\eta$ as a $d^2\theta$ integral. Then, using the same notation
as before, the soft terms take the  form
\eqn\general{{\cal L}_{soft}=\int d^2\theta \bigl\{\psi^2+\psi\fil^2+
(\psi +\psi\fil+\psi\fil^2+\psi^2)\eta+\dots\bigr\}}
were  $M_G\sim 1$. Again we can proceed as before, and redefine $\psi
\to \psi-(\eta+\fil^2+\fil\eta+\fil^2\eta)$. After which the
integration of $\psi$ and $V$ are clearly independent and go through
along the same lines as before. In particular
we can cast the contribution from the $V$ integration in the form
\softgauge. This shows that even with general
soft terms the only universal gauge contributions are represented by
D-terms. This result, however, can be important only in particular
models. In fact, as we show below, in the most general situation
the decoupling of the heavy chiral sector can in principle lead to
completely split soft terms.
In passing, we remind one well
known potential problem of the case of general soft terms, which
is that of the hierachy stability
\sen \barbieri\hitoshi. In order to maintain the hierarchy after the above
field redefinition the term $\fil^2\eta$ must be absent from the
superpotential. The conditions for this to happen are again rather
model dependent, though in particular models
a symmetry might be at the basis.

Let us now comment on the wide possibility of mass splittings
offered by the general case. Consider just
the following terms in the lagrangian
\eqn\split{\int d^2\theta \left\{ {1\over 2} (\mu_1)_{kl}\psi_l\psi_k+
(\mu_2)_{k\alpha}\psi_k\fil_\alpha\eta\right\}}
where $\mu_{1,2}={\cal O}(M_G)$ and are in general only $H$-symmetric.
Upon integrating out $\psi$ we get the soft mass term
\eqn\lightsplit{\int d^4\theta
\fild\mu_2^\dagger(\mu_1\mu_1^\dagger)^{-1}\mu_2\fil \eta\etab}
which is also in general only $H$-symmetric! In the aligned case
discussed in sect. 1,  not only $(\mu_2)_{k\alpha}
=(\mu_1\G)_{k\alpha}$,
which already typically implies a bigger symmetry in \lightsplit, but
there  is also an additional
 term $-|\G_{k\alpha}\fil_\alpha|^2$ which exactly
cancels the one above (see the second term in eq. \quadratic; we are
assuming the broken Cartan generators to be diagonal on the
light states). The implications of eq. \lightsplit\ are particularly
important in a scenario like $SO(10)$ Yukawa unification \anantha\raby,
were reducing the symmetry of the soft terms helps making electroweak
breaking more plausible \rs. Indeed, indicating
by $\underline {10}_1$ the multiplet containing the MSSM Higgs doublets, we
might have a mixing term $\underline{10}_1(M+\underline{45}_X)
\underline{10}_2\eta$, where $<\underline{45}_X>\propto X$ and
$\underline {10}_2$
has a direct GUT-scale mass $M_2\underline {10}_2^2$.
Eqs. \split\lightsplit\ then give the soft masses
$m_{H_i}^2=m^2|M+v_{45}
X_i|^2/M_2^2$, which split the light fields.
In a similar way we can imagine of coupling the matter representations
to some heavy vector $\underline {16}\oplus \underline{\bar {16}}$.
Then we can get ``vertically'' split masses of the form
$|a+b(B-L)+cT_{3R}|^2$,
where $a$, $b$ and $c$ are complex so that there are 5 free parameters which
can completely split the $\tilde Q,\tilde U_c,\tilde D_c,\tilde L,\tilde
E_c$ within a family. In this way we can get
 very general soft masses, the predictivity on the Yukawas notwithstanding.
Of course, in particular models, the terms we are describing could be
absent by the same reason that renders the MSSM fields light.
 For instance, one might expect that $\underline {10}_1$ couples to
heavy fields only via $\underline{10}_1(\underline{45}_{B-L})
\underline{10}_2\eta$ which reproduces
the superpotential term implementing the ``Dimopoulos-Wilczek''  mechanism
of doublet-triplet splitting \missing\babu\ ($<\underline {45}_{B-L}>
\propto B-L$). But this is not necessary. Even when a global symmetry
is  responsible for the specific form of the original
superpotential (like in ref. \raby), it is well possible
that the susy breaking spurion $\eta$ itself, representing now the
v.e.v. of a field, tranforms under the same symmetry. A new class of
soft terms is then allowed. On the other hand, what is really striking
about the aligned soft
terms of eq. (3), is that, even by allowing light-heavy mixings that
 individually look like the ones discussed in this section, the final result
eqs. \softgauge\softrest\  still bears a rather good memory of the original
gauge symmetry.

\vskip0.5truecm
\noindent{\bf 4.}{\hskip 0.2truecm} We have presented a rather
compact way of deriving the
tree level effective lagrangian below the GUT scale in softly
broken supersymmetry. We  focused first on soft terms coming from a
{\it hidden sector}.  We stressed how, in the superfield formalism,
supergauge invariance plays an important role in leading to the final
result. This is conveniently written in terms of the matrix
$\tG=\G-c_AT_A$ and of the so called D-terms. Its form is rather
constrained in cases of interest. We discussed the splittings of
sparticles embedded in a single GUT multiplet. In $SO(10)$ and $SU(5)$
these respect the $SU(5)$ symmetry, where in $SO(10)$ the reduction to
$SU(5)$ is determined just by
a universal D-term. In $SO(10)$ a further class of splitting effects
is forbidden by the requirement of $R$-parity conservation. However
effects in this class can become important  in $E_6$, due to
 the possible existence of
R-odd heavy vector superfields. These can lead to complete splitting of the
sparticles within one family. This is an interesting
fact that  had not been noticed before. We also have pointed out that
a similar result may hold, independent of the unified gauge
group, in a situation were the most general soft terms appear. This would
suggest a non-minimal scenario for supersymmetry breaking.
These last two remarks  on intrafamily splittings
could be very important for scenarios, like $SO(10)$ Yukawa
unification,  which are
otherwise rather constrained by their low-energy implications. They
 also  further confirm that the study of sparticle
spectroscopy will be of crucial help in selecting among various
scenarios for physics close to the Planck scale.

\vskip 1.0truecm

It is a pleasure to acknowledge stimulating discussions with Riccardo
Barbieri, Glennys Farrar, Dan Kabat, Hitoshi Murayama, Alex Pomarol,
Uri Sarid, Scott Thomas and
Fabio Zwirner. This
work was supported by the National Science Foundation under grant
PHY-91-21039.

\if\norefs n\listrefs\fi
\end